\documentclass[%
reprint,
superscriptaddress,
 amsmath,amssymb,
 aps,
 prx,
floatfix,
]{revtex4-2}

\usepackage{xcolor}
\usepackage{xfrac}

\newcommand{\singletS}{$^1S_0\,$}
\newcommand{\tripletP}{$^3P_1\,$}

\newcommand{\tripletPzero}{$^3P_0\,$}
\newcommand{\tripletPtwo}{$^3P_2\,$}

\newcommand{\tripletS}{$^3S_1\,$}

\usepackage{graphicx}
\usepackage{dcolumn}
\usepackage{bm}
\usepackage[section]{placeins}

\begin{document}

\title{Trapped arrays of alkaline earth Rydberg atoms in optical tweezers}

\author{J.T. Wilson}
    \thanks{These authors contributed equally to this work.}
\affiliation{Department of Electrical Engineering, Princeton University, Princeton, NJ 08540}

\author{S. Saskin}
    \thanks{These authors contributed equally to this work.}
\affiliation{Department of Electrical Engineering, Princeton University, Princeton, NJ 08540}
\affiliation{Department of Physics, Princeton University, Princeton, NJ 08540}

\author{Y. Meng}
\affiliation{Vienna Center for Quantum Science and Technology, TU Wien, Atominstitut, Stadionallee 2, 1020 Vienna, Austria}

\author{S. Ma}
\affiliation{Department of Electrical Engineering, Princeton University, Princeton, NJ 08540}
\affiliation{Department of Physics, Princeton University, Princeton, NJ 08540}

\author{R. Dilip}
\affiliation{Department of Physics, Princeton University, Princeton, NJ 08540}

\author{A.P. Burgers}
\affiliation{Department of Electrical Engineering, Princeton University, Princeton, NJ 08540}

\author{J.D. Thompson}
 \email{jdthompson@princeton.edu}
\affiliation{Department of Electrical Engineering, Princeton University, Princeton, NJ 08540}

\date{\today}

\begin{abstract}
Neutral atom qubits with Rydberg-mediated interactions are a leading platform for developing large-scale coherent quantum systems. In the majority of experiments to date, the Rydberg states are not trapped by the same potential that confines ground state atoms, resulting in atom loss and constraints on the achievable interaction time. In this work, we demonstrate that the Rydberg states of an alkaline earth atom, ytterbium, can be stably trapped by the same red-detuned optical tweezer that also confines the ground state, by leveraging the polarizability of the Yb$^+$ ion core. Using the previously unobserved \tripletS series, we demonstrate trapped Rydberg atom lifetimes exceeding $100\,\mu$s, and observe no evidence of auto- or photo-ionization from the trap light for these states. We measure a coherence time of $T_2 = 59$ $\mu$s between two Rydberg levels, exceeding the 28 $\mu$s lifetime of untrapped Rydberg atoms under the same conditions. These results are promising for extending the interaction time of Rydberg atom arrays for quantum simulation and computing, and are vital to capitalize on the extended Rydberg lifetimes in circular states or cryogenic environments.

\end{abstract}

\maketitle

Arrays of individually-trapped neutral atoms with strong interactions via Rydberg excitations are a promising platform for quantum simulation, optimization and computing \cite{Lukin2001DipoleEnsembles,Saffman2010}. The combination of a flexible geometry and highly controllable interactions has enabled explorations of many-body quantum dynamics \cite{Bernien2017ProbingSimulator,Lienhard2018ObservingInteractions, deLeseleuc2019ObservationAtoms}, high-fidelity gates \cite{Urban2009ObservationAtoms,Wilk2010EntanglementBlockade,Jau2016EntanglingBlockade, Levine2018High-FidelityQubits,Levine2019ParallelAtoms, Graham2019} and the generation of extremely large entangled states \cite{Omran2019GenerationArrays}. The majority of existing work uses alkali atoms, but recent experiments with ground state alkaline earth atoms in optical tweezers \cite{Cooper2018Alkaline-EarthTweezers, Norcia2018MicroscopicArrays, Saskin2019} suggest a number of technical advantages as well as the potential to apply entangled states to enhance optical atomic clock performance \cite{Norcia2019Seconds-scaleArray, Madjarov2019AnReadout}.

A central challenge to experiments with Rydberg atoms in standard, red-detuned optical tweezers is that they are repelled from the intensity maximum. This repulsion arises from the ponderomotive potential of the essentially free Rydberg electron, described by the polarizability $\alpha_p = -e^2/m_e \omega^2$, which is always negative \cite{Dutta2000} (here, $\omega$ denotes the frequency of the trap light, and $e$, $m_e$ are the electron charge and mass). To mitigate this effect, the vast majority of experiments operate with the tweezers turned off during the Rydberg excitation, which limits the interaction time to 10-20 $\mu$s because of the expansion of the atoms at typical temperatures of 10-20 $\mu$K. This is significantly below the typical room temperature Rydberg state lifetime of 100-300 $\mu$s for $n=60-100$ $S$ states \cite{Archimi2019MeasurementsLevels, Saffman2010}, and far below the tens of seconds achievable with circular states in cryogenic cavities \cite{Nguyen2018}. Furthermore, heating associated with modulating the trap may impact the gate fidelity in sequential operations.

In recent work, it has been demonstrated that the ponderomotive potential can be used to trap Rydberg atoms in a 3D intensity minimum. Rubidium Rydberg states have been trapped for up to 200 $\mu$s in a hollow ``bottle beam" generated by a spatial light modulator \cite{Barredo2019}, while simultaneous trapping of ground and Rydberg states has been achieved in a lattice of blue-detuned light sheets, with 50 $\mu$s dwell time for atoms in Rydberg states \cite{Graham2019}. The stability of these traps requires that the spatial extent of the intensity minimum is large compared to the Rydberg electron orbit ($R_e = 3n^2 a_0/2 \approx 0.8\,\mu$m for $n=100$). This necessitates a large-waist optical trap, a corresponding increase in total optical power per trap, and imposes a maximum principal quantum number that can be trapped for a given power, of order $n=90$ in Ref. \cite{Barredo2019}. Ensembles of Rydberg atoms have also been trapped using several approaches \cite{Anderson2011,Li2013EntanglementExcitation,Anderson2013,Hogan2008,Cortinas2019LaserAtoms}.

\begin{figure}[htp]
\includegraphics{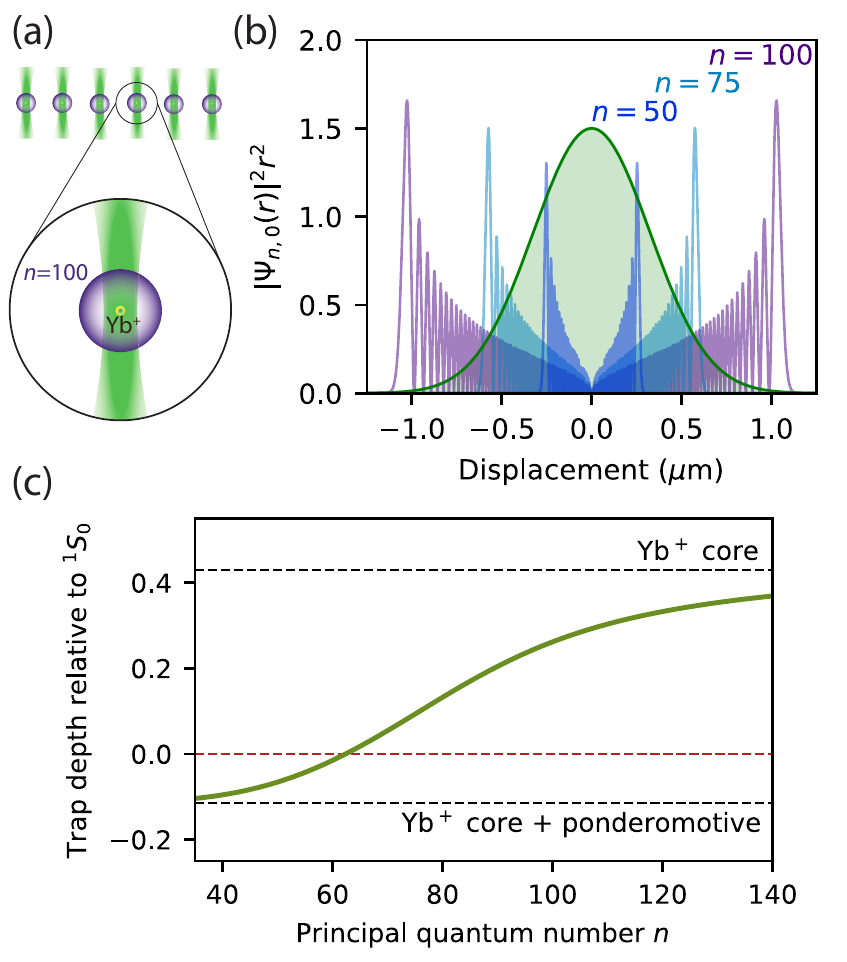}
\caption{\label{fig:introfig}(a) Cartoon of the experiment, showing a six-tweezer array, the Rydberg electron wavefunction, and the Yb$^+$ ion core. (b) Radial probability distributions of Rydberg wavefunctions relative to the optical tweezers (green). (c) Calculated trap depth for \tripletS states, normalized to the \singletS ground state for the same power and beam waist (here, $0.65\,\mu$m).
}
\end{figure}

In this work, we demonstrate an alternate approach: leveraging the polarizability of the Yb$^+$ ion core to directly trap Yb Rydberg atoms in conventional, red-detuned optical tweezers \cite{Mukherjee2011,Topcu2014}. Unlike alkali atoms, the ion core of alkaline earth atom Rydberg states has significant polarizability at typical laser trapping wavelengths. The ponderomotive potential of the Rydberg electron contributes an anti-trapping effect, but it is small for short wavelengths and high-$n$ Rydberg states where the beam waist is comparable to or smaller than $R_e$. We demonstrate trap lifetimes exceeding 100 $\mu$s for $n=75$ with less than 10 mW of optical power per trap. Trap-induced losses from photo-ionization are negligible for $S$ states, but slightly shorten the lifetime of $P$ and $D$ states. We study the interplay of the ponderomotive and Yb$^+$ core potentials in detail, including the dependence on the Rydberg level, and observe that ``magic" trapping is possible for certain pairs of Rydberg states. A theoretical model is presented to efficiently calculate the trapping potentials based on a decomposition of the optical tweezer into irreducible tensor operators. We study the coherence properties of a superposition of trapped Rydberg levels, achieving $T_2 = 59$ $\mu$s, limited by finite temperature and the differential light shift of the two states in the trap but exceeding the lifetime of the Rydberg atom in the absence of the trap. This work also presents the first measurement of the lifetime of high-$n$ Yb Rydberg states and the first observation of the \tripletS Yb Rydberg series.

The trapping potential for Ytterbium Rydberg states with the configuration $6snl$ arises from separate contributions from the $6s$ core and $nl$ Rydberg electrons \cite{Topcu2014}. In SI units, the core potential $U_c(\vec{R}) = - \frac{1}{2 \epsilon_0 c} \alpha_c(\omega) I(\vec{R}) $ is derived from the dynamic electric dipole polarizability $\alpha_c(\omega)$ of the Yb$^+$ ion $6s$ $^2S_{1/2}$ state (here, $I(\vec{R})$ is the light intensity at the nuclear coordinate $\vec{R}$, $\epsilon_0$ is the permittivity of free space, and $c$ is the speed of light). For the 532 nm light used here, this is of the same order of magnitude as the Yb$^0$ ground state potential, as the principal Yb$^+$ transitions (369, 329 nm) are not too far from the principal Yb$^0$ transition (399 nm). The nearly free Rydberg electron experiences a ponderomotive potential that depends on the intensity averaged over its wavefunction \cite{Dutta2000}:
\begin{equation}
\label{eq:upond}
U_r(\vec{R}) = \frac{e^2}{2 \epsilon_0 c m_e \omega^2} \int \left| \psi_{nl}(\vec{r})\right|^2 I(\vec{r}+\vec{R}) d^3\vec{r}.
\end{equation}
Here, $\psi_{nl}(\vec{r})$ is the wavefunction of the $nl$ electron ($\vec{r}$ is the electron coordinate relative to the nucleus; Fig. \ref{fig:introfig}b). In Fig. \ref{fig:introfig}c, the sum of these contributions for the \tripletS Rydberg states in an optical tweezer ($\lambda = 532$ nm, $1/e^2$ radius $w_0 = 650$ nm) is shown as a function of the principal quantum number $n$. For low $n$ where the Rydberg wavefunction is significantly smaller than the beam waist, the total polarizability is $\alpha_c(\omega) - e^2/m \omega^2$, while at high $n$ it asymptotes to $\alpha_c(\omega)$, as the overlap of the Rydberg electron with the tweezer decreases.

We characterize the trapping potential for Yb Rydberg states using an array of six optical tweezers loaded with single $^{174}$Yb atoms as described previously in Ref. \cite{Saskin2019}. A large array spacing ($d=24\,\mu$m) minimizes the influence of interactions on the spectroscopy. We excite atoms to Rydberg states using sequential single-photon $\pi$ pulses on the \singletS $\rightarrow$ \tripletP $(M_J=-1)$ and \tripletP $(M_J=-1) \rightarrow 6sns$\,\tripletS $(M_J=-1)$ transitions; this configuration is somewhat inefficient because of the finite lifetime of the intermediate state (860 ns), but avoids noise on our 556 nm laser system that was not designed for coherent two-photon excitation. The 308 nm light for the Rydberg transition is generated by summing a Ti:Sapphire (TiS) laser with a 1565 nm fiber laser and doubling the 616 nm output in a resonant cavity. We have generated more than 100 mW in this configuration, but the experiments described here used approximately 5 mW focused to 10 $\mu$m. We primarily study the Yb 6sns \tripletS series, which has not been previously observed to the best of our knowledge. The series is relatively unperturbed, with a quantum defect of approximately 4.439 (additional details are provided in the supplementary information \cite{supplement}). For this state, we achieve a Rabi frequency of $\Omega = 2\pi \times 2.5$ MHz from \tripletP.

\begin{figure}[htp]
\includegraphics{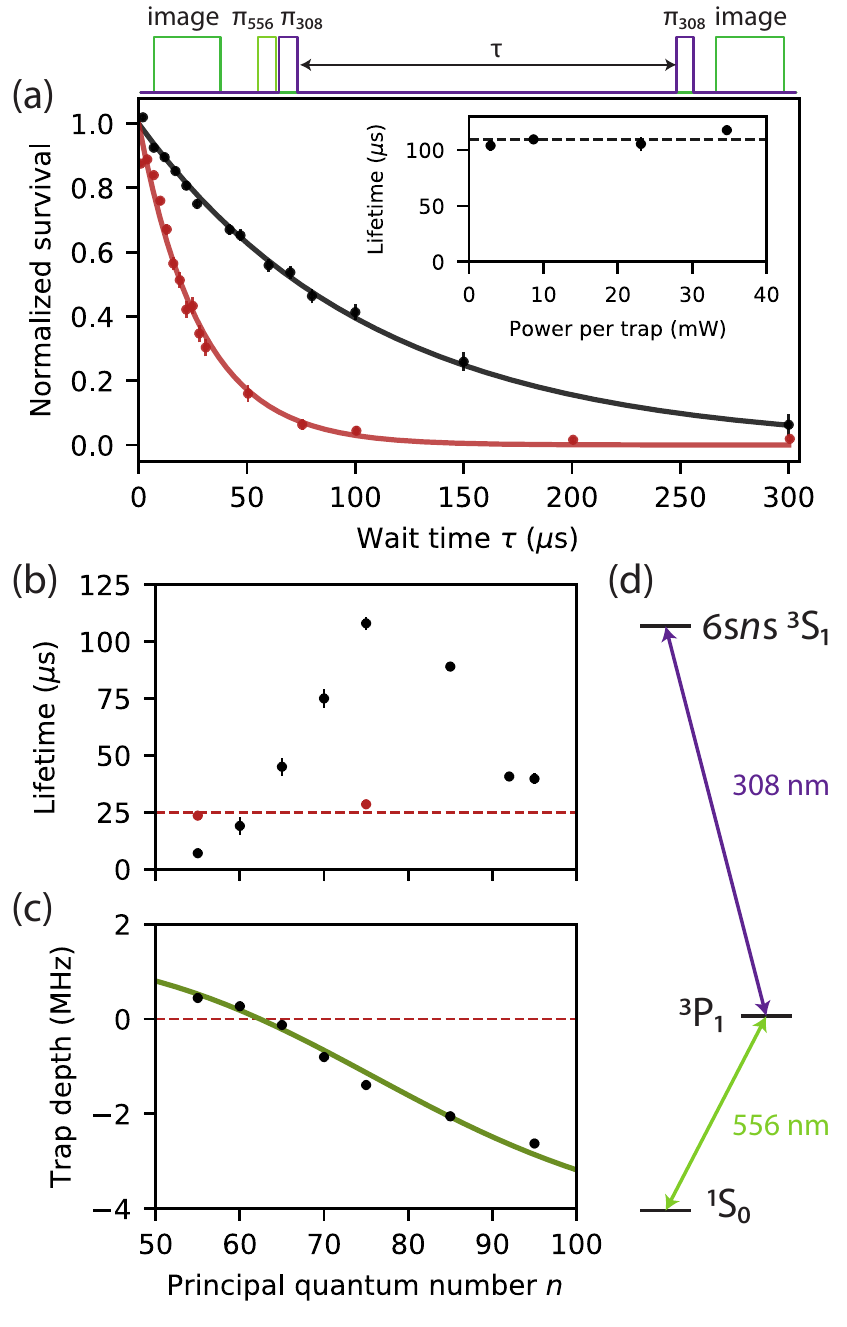}
\caption{\label{fig:fig2}(a) Survival probability of the $n=75$ \tripletS state with (black, $\tau = 108\;\mu$s) and without (red, $\tau = 28\;\mu$s) the traps. Inset: Trapped Rydberg lifetime $\tau$ of the \tripletS Rydberg state vs. trap power at $n=75$. (b) Trapped Rydberg lifetime of the \tripletS state vs. principal quantum number $n$, with (black) and without (red) the trap. The dashed red line shows the untrapped lifetime of a ground state atom under the same conditions. (c) Trap depth of the \tripletS Rydberg state vs principal quantum number. The green line is the theoretical trap depth using the calculation from Fig. \ref{fig:introfig}c. (d) Relevant Yb energy levels for Rydberg excitation.}
\end{figure}

We measure the trapped lifetime of a Rydberg atom by imaging the array of ground state atoms, exciting to a particular Rydberg state, waiting a variable time $\tau$, and de-exciting using a second UV laser pulse before acquiring a second image. If the Rydberg atom leaves the trap or changes states between the UV pulses (\emph{i.e.} from spontaneous decay or interaction with blackbody radiation), it will not be de-excited by the second pulse and will be recorded as an atom loss between the two images. A typical trace for the $n=75$ \tripletS state is shown in Fig. \ref{fig:fig2}a using 9 mW per trap (12 MHz ground state trap depth). If the trap is turned off between the UV pulses, the Rydberg atom survives for 28 $\mu$s, consistent with the measured ground state lifetime in the absence of a trap. When the trap is on between the UV pulses, the lifetime is extended to 108 $\mu$s. To investigate the role of trap-induced loss processes such as photo- or auto-ionization of the Rydberg state, we measure the lifetime as a function of the trap depth, shown in the Fig. \ref{fig:fig2}a inset. We observe no influence of the trap depth on the lifetime over a wide range of powers.

We repeat these measurements at several values of principal quantum number $n$. At low $n$ (e.g. $n=55$), the lifetime with the trap is shorter than without the trap, suggesting that these states are repelled. Above $n \approx 60$, the trapped lifetimes are longer, consistent with trapping. Curiously, they reach a maximum at $n=75$ and then decrease, although the intrinsic Rydberg lifetimes are expected to increase monotonically as $n^2$. We do not observe any trap power dependence of the lifetime between $n=70$ and $n=95$, ruling out trap-induced losses. We conjecture that noise or cavity effects from our in-vacuum electrodes may play a role in the reduction of the lifetime \cite{supplement}. 

To study the interplay of the ponderomotive and core ion polarizabilities, we measure the trap depth as a function of $n$ using the AC stark shift of the UV \tripletP to \tripletS transition. We measure a crossover from anti-trapping to trapping around $n=60$, consistent with the onset of the lifetime increase. To obtain the absolute shift of the Rydberg state in the trap, we subtract the \tripletP trap depth, which we infer from the measured \tripletP-\singletS light shift in the trap ($7.54$ MHz) and the ratio of the polarizabilities of these states $R = \alpha_{\;^3P_1}/\alpha_{\;^1S_0} \approx 0.39$  \cite{Yamamoto2016AnCooling}. Because of uncertainty in $R$, there is a systematic uncertainty of $\sim 0.2$ MHz in the Rydberg trap depth, which allows the crossover $n$ between trapping and anti-trapping to vary between $56$ and $63$. Fixing it at $n=62$ gives good agreement with a model with $w_0=650$ nm and $\alpha_c(532 \; \textrm{nm}) = 107$ a.u., within 12\% of the value calculated in Ref. \cite{Roy2017}.

\begin{figure}[htp]
\includegraphics{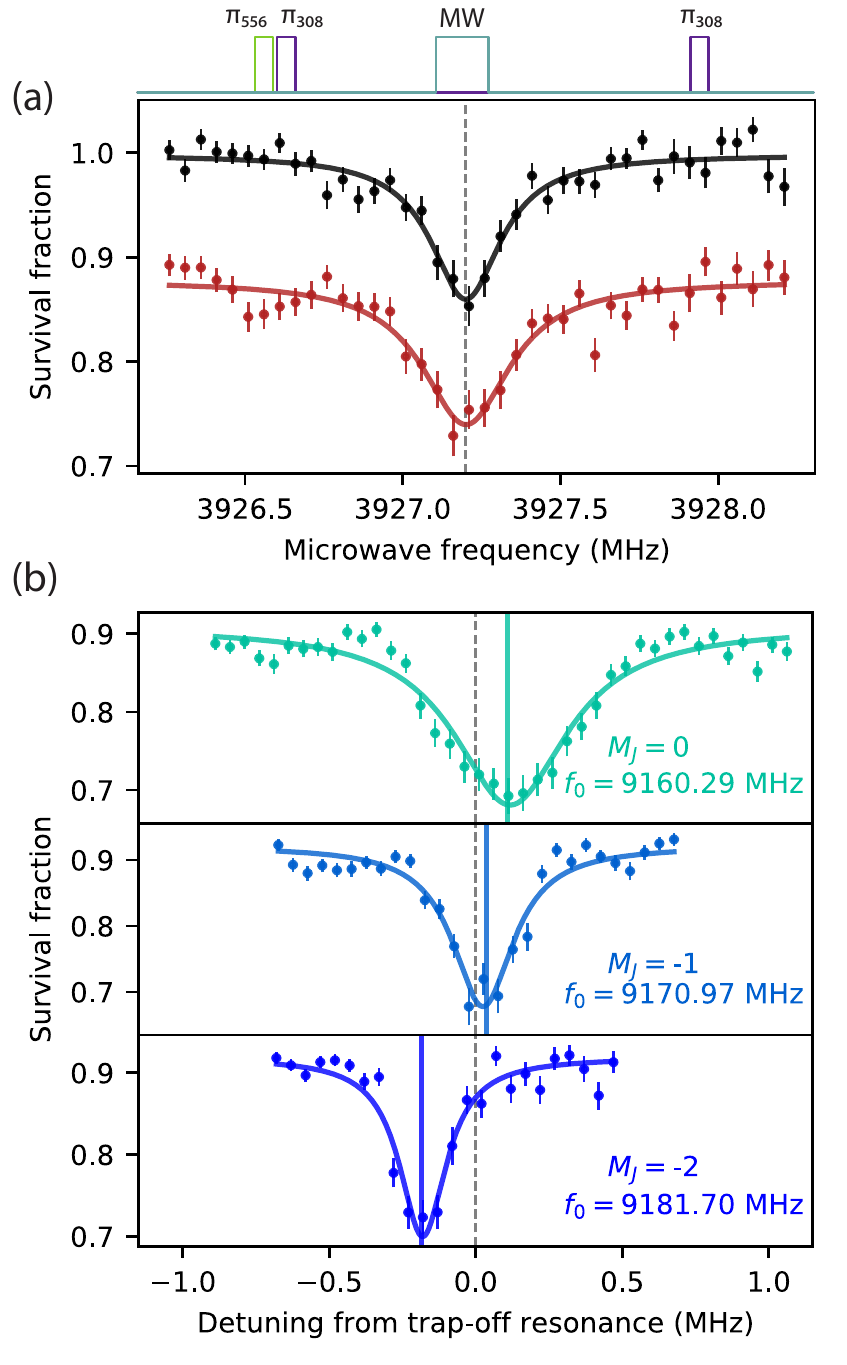}
\caption{\label{fig:fig3}(a) Microwave spectrum of the $n=75$ \tripletS to $n=74$ \tripletPzero transition with (black) and without (red) the traps, demonstrating the magic trapping condition. The black data are shifted for clarity and the solid lines are Lorentzian fits. (b) Microwave spectra of the n=75 \tripletS $M_J=-1$ to $n=74$ \tripletPtwo $M_J=-2,-1,0$ transitions, showing the tensor light shift of different $M_J$ levels from the ponderomotive potential. For each transition, zero detuning indicates the measured transition frequency without the trap, indicated in the figure. The solid vertical lines show the predicted $M_J^2$ dependence of the tensor light shift.}
\end{figure}

Next we study the state-dependent nature of the trapping potential by driving microwave transitions between Rydberg states following optical excitation to a \tripletS state (Fig. \ref{fig:fig3}). The shift of the microwave transition when the dipole trap is applied probes the differential polarizability of these states. The \tripletS and \tripletPzero states have nearly vanishing differential polarizability: on top of an estimated trap depth of $1.4$ MHz, the transition frequency shifts less than 10 kHz. This is in agreement with a theoretical prediction \cite{supplement} that the \singletS, \tripletS, and \tripletPzero states should experience the same, purely scalar, ponderomotive potential, and the fact that the ion core polarizability is independent of the state of the Rydberg electron. In contrast, the \tripletPtwo state has a strong $M_J$-dependent shift arising from the rank-2 (tensor) component of the ponderomotive potential (Fig. \ref{fig:fig3}b). Intuitively, this is results from the different orientations of the $M_J$ angular wavefunctions with respect to the tweezer potential, which is not spherically symmetric. The observed tensor shift of $300$ kHz is close to the computed value of $400$ kHz using the model parameters discussed above.

We have also measured the lifetimes of several $P$ and $D$ states, presented in the supplementary information \cite{supplement}. Near $n=75$, the $^3P_2$ and $^1D_2$ lifetimes are similar to $^3S_1$, while the $^3P_0$ lifetime is nearly 10 times shorter, presumably because this series is very strongly perturbed \cite{Aymar1984Three-stepYtterbium}. However, both $P$ and $D$ states experience a moderate reduction in lifetime with increasing trap power, attributable to photo-ionization. The approximate magnitude and $L$-dependence are in approximate agreement with previous calculations for Rb \cite{Saffman2005}.

\begin{figure*}[htp]
\includegraphics{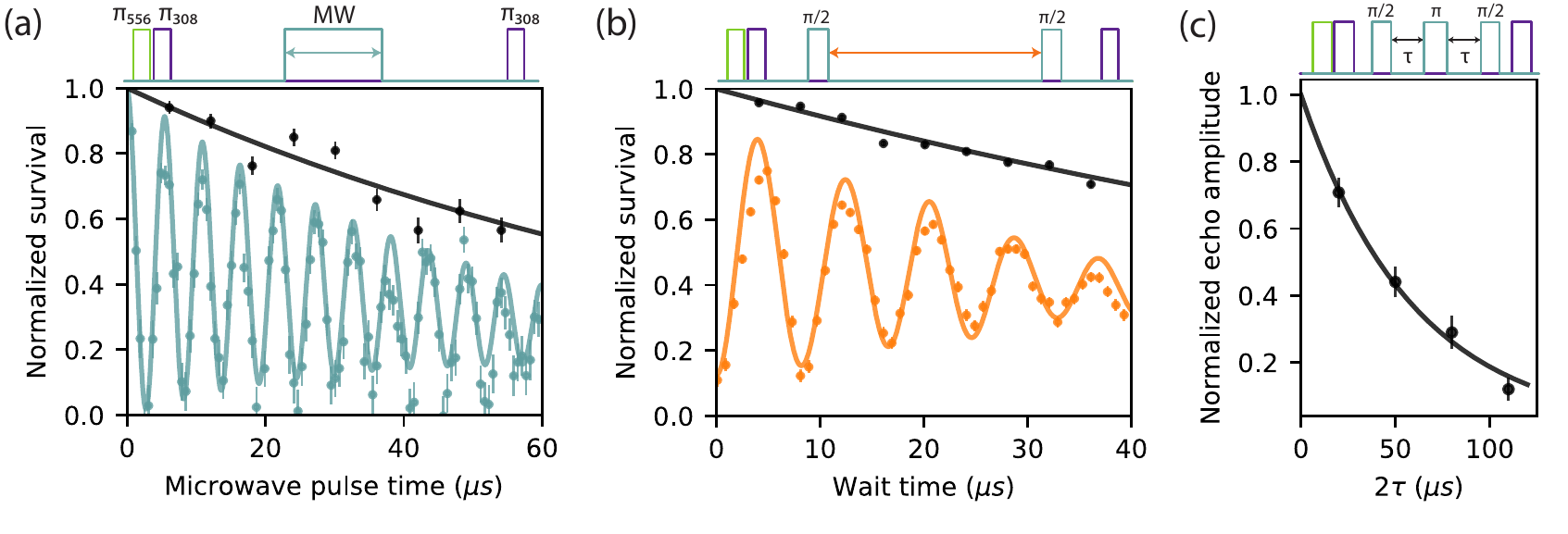}
\caption{\label{fig:fig4}(a) Two-photon Rabi oscillations between $n=75$ \tripletS and $n=74$ \tripletS. The solid line is a cosine fit with exponential decay time $\tau=42\,\mu$s. Control data without microwave pulses (black data) shows $T_1$ for comparison. (b) Ramsey measurement of $T_2^*$. The orange line is a simulation that takes into account dephasing from the differential light shift between the two levels (90 kHz) and a finite atomic temperature (13 $\mu$K), yielding a $1/e$ decay time of $22\,\mu$s. (c) Hahn echo measurement. The black line is an exponential fit that yields $T_2 = 59\,\mu$s.}
\end{figure*}

To demonstrate the utility of trapping Rydberg states for quantum simulation and quantum computing, we probe the coherence properties of a superposition of Rydberg levels. In Fig. \ref{fig:fig4}a, we show Rabi oscillations between the $n=74$ and $n=75$ \tripletS states, driven by a two-photon microwave transition detuned by 40 MHz from the \tripletPzero intermediate state. The oscillations persist for more than 60 $\mu$s, more than twice the lifetime of an untrapped Rydberg atom. The coherence time is quantified using a Ramsey sequence (Fig. \ref{fig:fig4}b) and found to be $T_2^* = 22\,\mu$s, which is in agreement with dephasing from thermal motion \cite{Kuhr2005AnalysisTrap} for an atom with a temperature of $T=13\ \mu$K and the (measured) difference in the potential depth for the two states of $90$ kHz. A Hahn echo sequence yields $T_2 = 59\,\mu$s. This is shorter than the limit $T_2 = T_1$ because the axial trap period is longer than $T_1$, so the revival in coherence is not reached (here, $T_1 = 108\,\mu$s is the lifetime of the upper \emph{and} lower states).

These results demonstrate that trapping Rydberg states of alkaline earth atoms using the core polarizability can extend the coherence of quantum operations beyond what is possible with un-trapped atoms. This may lead to improved fidelities for quantum simulators and Rydberg gates, especially between distant atoms when the interaction is small. The expected improvement from trapping Rydberg states is most significant when the Rydberg lifetimes are very long, as expected for low-$l$ states at cryogenic temperatures, and especially circular Rydberg states.

We conclude with a discussion of several aspects of these results. First, the coherence times in Fig. \ref{fig:fig4} are limited by a slight $n$-dependence of the trapping potential. While the ponderomotive potential itself is only weakly $n$-dependent, it contributes a large fractional $n$-dependence when it is almost completely cancelled by the $n$-independent core potential near the crossover at $n\approx 60$. In future experiments where a higher degree of state-insensitive trapping is required, this can be improved by using higher $n$ states or by pushing the crossover to lower $n$ using shorter wavelength trapping light (to increase the core polarizability and decrease the ponderomotive potential) or smaller beam waist. Tuning the beam waist allows the precise potential for a particular Rydberg state to be manipulated, which may be advantageous for fine-tuning triply-magic trapping of ground, clock and Rydberg states, for example \cite{Topcu2016PossibilityLattices}.

Second, we consider the prospect of trapping circular Rydberg states of Yb, which is particularly intriguing as these states have been predicted to have lifetimes of tens of seconds in cryogenic microwave cavities \cite{Nguyen2018}. The photoionization rate of the $P$ and $D$ states shortens their lifetime from their intrinsic values by 15-30\% at a trap power of 9 mW per tweezer. However, for high-$L$ states, including circular states, the photoionization cross-section is significantly smaller, ultimately decreasing exponentially with $L$ \cite{Nguyen2018}. The auto-ionization rate, while immeasurably small in our current experiments, should decrease as $L^{-5}$ as the contact of the Rydberg electron with the core is suppressed \cite{Cooke1979CalculationStates}. Therefore, it seems likely that circular states may be trapped for extremely long times without adverse affects. Furthermore, transfer of orbital angular momentum from focused Laguerre-Gauss modes through the ponderomotive potential offers an intriguing new route to efficiently and rapidly exciting circular Rydberg states \cite{Cardman2019CircularizingTraps} or driving transitions between them.

We gratefully acknowledge Tom Gallagher and Patrick Cheinet for helpful conversations about Yb Rydberg states, Shimon Kolkowitz for discussions about the manuscript, and Toptica Photonics for the loan of a WS8-2 wavemeter used for the Rydberg spectroscopy. This work was supported by the Army Research Office (contract W911NF-18-1-0215) and the Sloan Foundation. J.W. was supported by an NSF GRFP. S.S. was supported by an ARO QuaGCR fellowship. Y.M. was supported by the Austrian Science Fund (DK CoQuS Project No. W 1210-N16).

\FloatBarrier
\bibliography{finalbib}
\bibliographystyle{apsrev4-2}
\end{document}


\title{Supplementary Information for ``Trapped arrays of alkaline earth Rydberg atoms in optical tweezers"}

\author{J.T. Wilson}
    \thanks{These authors contributed equally to this work.}
\affiliation{Department of Electrical Engineering, Princeton University, Princeton, NJ 08540}

\author{S. Saskin}
    \thanks{These authors contributed equally to this work.}
\affiliation{Department of Electrical Engineering, Princeton University, Princeton, NJ 08540}
\affiliation{Department of Physics, Princeton University, Princeton, NJ 08540}

\author{Y. Meng}
\affiliation{Vienna Center for Quantum Science and Technology, TU Wien, Atominstitut, Stadionallee 2, 1020 Vienna, Austria}

\author{S. Ma}
\affiliation{Department of Electrical Engineering, Princeton University, Princeton, NJ 08540}
\affiliation{Department of Physics, Princeton University, Princeton, NJ 08540}

\author{R. Dilip}
\affiliation{Department of Physics, Princeton University, Princeton, NJ 08540}

\author{A.P. Burgers}
\affiliation{Department of Electrical Engineering, Princeton University, Princeton, NJ 08540}

\author{J.D. Thompson}
 \email{jdthompson@princeton.edu}
\affiliation{Department of Electrical Engineering, Princeton University, Princeton, NJ 08540}

\date{\today}

\maketitle

\section{Calculation of Rydberg Trapping Potentials}
\label{trapping calc}

To calculate the potential seen by a Rydberg atom in an optical tweezer, we compute the energy as a function of the position of the atomic nucleus $\vec{R}$. To make the calculation tractable, we restrict our calculation to states with electron configuration $6snl$, and make the approximation that the inner $6s$ and outer $nl$ electrons can be treated separately \cite{Topcu2014}. 

The inner electron polarizability is dominated by the Yb$^+$ ion $6s-6p$ transitions. For linearly polarized trapping light, there is only a scalar polarizability, which has been computed at 532 nm to be $\alpha_c = 96$ atomic units (a.u.) for the $^2 S_{1/2}$ ground state \cite{Roy2017}, giving rise to a potential $U_{c} = -\frac{1}{2 \epsilon_0 c}\alpha_c I$. $\alpha_c$ is around 35\% of the Yb$^0$ \singletS polarizability (275 a.u. \cite{Guo2010} or 226 a.u. \cite{Dzuba2010}), indicating that suitably deep traps can be reached with the same powers used to trap ground states. Intriguingly, it is also very close to the calculated polarizability of the metastable Yb$^0$ $6s6p$ \tripletPzero level (95 a.u. \cite{Guo2010} or 91 a.u. \cite{Dzuba2010}), which may enable magic-wavelength trapping \cite{Topcu2014} of high-$n$ Rydberg states and the upper clock state.

The outer, Rydberg electron experiences a spatially-dependent ponderomotive potential proportional to the intensity of the trapping light:

\begin{equation}
    U_{r} = -\frac{1}{2 \epsilon_0 c} \alpha_p \int d^3\vec{r} |\psi(\vec{r})|^2 I(\vec{r}+\vec{R}).
\end{equation}
Here, $\alpha_p = -e^2/m_e \omega^2$ is the ponderomotive polarizability, $\vec{r}$ is the coordinate of the electron with respect to the nucleus at $\vec{R}$, and $\psi(\vec{r})$ is the electronic wavefunction.  This integral is the expectation value of the operator $I$ for the electron wavefunction. Its evaluation is simplified if $I(r)$ can be expanded in irreducible tensor operators; then, the Wigner-Eckart theorem allows $U_{r}$ to be expressed for any state in terms of angular factors and fewer than $L+1$ radial integrals. For a lattice, this expansion can be done analytically \cite{Topcu2013,Knuffman2007}; however, for a tweezer or other, arbitrary potential it must be done numerically. Specifically, we seek an expansion in spherical harmonics centered on the nuclear coordinate $\vec{R}$:

\begin{equation}
    I(\vec{r}+\vec{R}) = \sum_{kq} I^{(k)}_q = \sum_{kq} f_{kq}(r;\vec{R}) C^{(k)}_q (\hat{r}).
\end{equation}
Here, $C^{(k)}_q (\hat{r}) = \sqrt{\frac{4 \pi}{2 k+1}} Y^{(k)}_q(\hat{r})$ is the normalized spherical harmonic and $r = |\vec{r}|$ and $\hat{r} = \vec{r}/r$, and $f_{kq}(r)$ are the coefficients to be found representing $I$. These functions, which have $\vec{R}$ as a parameter, can be computed by exploiting the orthonormality of the spherical harmonics as:

\begin{equation}
    f_{kq}(r;\vec{R}) = \sqrt{\frac{2 k+1}{4 \pi}} \iint d\Omega I(\vec{R}+\vec{r}) Y^{(k)}_q(\theta,\phi).
\end{equation}
Here, the angular integration is performed with respect to $\vec{r}$.

With the potential decomposed in this way, we can evaluate the matrix element between arbitrary states with quantum numbers $n'l'm'$ and $nlm$ using the Wigner-Eckart theorem:

\begin{equation}
    \label{eq:upond_nlm}
    \langle n'l'm'| I | nlm \rangle = \sum_{kq} \langle n'l'm'| I^{(k)}_q | nlm \rangle = \\
    \sum_{kq} (-1)^{l'-m'} \threej{l'}{k}{l}{-m'}{q}{m} \langle n'l' || I^{(k)}_q || nl \rangle.
\end{equation}
The reduced matrix element is given by

\begin{equation}
    \label{eq:reducednl}
    \langle n'l' || I^{(k)}_q || nl \rangle =\\ (-1)^{l'} \sqrt{(2l'+1)(2l+1)}  \threej{l'}{k}{l}{0}{0}{0}\\ \int dr r^2 R_{n'l'}(r)R_{nl}(r) f_{kq}(r; \vec{R}),
\end{equation}
Here, the term in parentheses is a Wigner $3j$ symbol, and $R_{nl}$ is the radial wavefunction of the Rydberg electron. The potential for a state $nlm$ is:
\begin{equation}
    \label{eq:Ur}
    U_{r} = -\frac{1}{2 \epsilon_0 c} \alpha_p \langle nlm| I | nlm \rangle.
\end{equation}

This calculation describes the action of $U_{r}$ in the $nlm$ basis, but the Rydberg states of alkali and alkaline atoms have significantly resolved fine structure splittings. Therefore, these expressions must be expanded in the appropriate basis. Before proceeding further, we note several properties of the potential that are already apparent. First, the contribution of odd-$k$ terms vanishes between states of the same $l$. Second, $f_{kq} = 0$ when $q \neq 0$ for $\vec{R}$ on the $z$-axis, since the potential is cylindrically symmetric. Lastly, only terms with $k \leq l$ have non-vanishing contributions, in order to satisfy the conservation of angular momentum (here, $l$ is the total angular momentum, but in the fine structure basis, this will be replaced by $j = l+s$).

Taken together, these allow the potential $U_{r}$ for low-$l$ states at $\vec{R}=0$ to be evaluated from a small number of $f_{kq}$, with $k$ even and $q=0$. At high-$l$ (i.e., circular states) it appears that a large number of $f_{kq}$ contribute to the potential. However, the radial matrix elements and angular coefficients of high-$l$ states decay rapidly with $k$, fundamentally as a consequence of the fact that a gaussian beam does not have significant angular momentum, and we find that truncating the calculation to the lowest few $k$ is an excellent approximation.

Additionally, we note that some of the off-diagonal matrix elements $\langle n'l'm'| I | nlm \rangle$ are of similar magnitude to the diagonal elements, and in this sense Eq. \ref{eq:Ur} is only the first order term in a perturbative calculation of the energy shift. For the trap depths considered here, the total ponderomotive potential is smaller than the spacing between all of the unperturbed energy levels (including $m$ levels, as we apply a $4.5$ G magnetic field), such that the next order terms do not change the potential significantly. However, in the absence of external fields, the off-diagonal terms in $m$ can be significant away from the $z-$axis. More generally, the terms between different $l,m$ may be exploited to drive transitions between Rydberg states in an intensity-modulated beam, where the modulation is resonant with the energy difference between the initial and final states. This could be particularly useful in connection with high-order Laguerre-Gauss beams to create large $k$ terms to efficiently excite circular states (following Ref. \cite{Cardman2019CircularizingTraps}) or drive transitions between circular states. Unlike two-photon electric dipole transitions, which are limited to $k\leq2$, the ponderomotive potential can drive many angular momentum quanta in a single step.

Lastly, we note that the $\vec{R}$ dependence of $U_r$ is not the same as $I(\vec{R})$, because of spatial averaging by the extended Rydberg electron wavefunction. In this work, we consider only the trap depth $U_r(\infty) - U_r(0) = - U_r(0)$. From numerical evaluations of $U_r(\vec{R})$, we have observed that the trap frequencies are slightly higher than would be predicted from the depth alone, since the spatial averaging washes out the (repulsive) ponderomotive potential.

We now consider the evaluation of the reduced matrix elements separately for alkali atoms and alkaline earth atoms.

\subsection{Alkali atoms}
 To compute the potential for alkali atoms, the matrix elements are needed in the spin-orbit (fine structure) basis: $\langle nsljm | I | nsljm \rangle$. As before, we start with the Wigner-Eckart theorem:

\begin{equation}
\label{eq:wesljm}
    \langle nsljm | I^{(k)}_q | nsljm \rangle =\\ (-1)^{j-m} \threej{j}{k}{j}{-m}{q}{m} \langle nslj || I^{(k)}_q || nslj \rangle.
\end{equation}
Then, we reduce the matrix element again to one only acting on $l$ \cite{Edmonds1996}:

\begin{equation}
\label{eq:finalalkali}
    \langle nslj || I^{(k)}_q || nslj \rangle =\\ (-1)^{s+l+j+k} (2j+1) \sixj{l}{j}{s}{j}{l}{k} \langle nl || I^{(k)}_q || nl \rangle.
\end{equation}
The reduced matrix element here is the same as Eq. (\ref{eq:reducednl}), and the coefficients are tabulated in Table \ref{tab:angular}.

\subsection{Alkaline earth atoms}
In the case of divalent Yb, we are interested in calculating the trapping potential for Rydberg states with term symbols $^{2S+1}L_J$ and $6snl$ electronic configurations. The LS-coupled basis is a close approximation to the true eigenbasis: in Yb the measured single-triplet mixing arising from spin-orbit coupling in high-$n$ $^3P_1$ states is approximately 6\% \cite{Neukammer1984DiamagneticExtent}, and this effect is presumably smaller in lighter alkaline earth atoms such as Sr. The ponderomotive potential only acts on the outer electron, so we reduce the matrix elements to account for this. 

Using the Wigner-Eckart theorem, we calculate the diagonal matrix elements between LS-coupled states with $S,L,J$ denoting the total electronic spin, orbital angular momentum and overall angular momentum, and $M$ the z-projection of $J$:

\begin{equation}
    \label{eq:nSLJM}
    \langle nSLJM | I^{(k)}_q | nSLJM \rangle =\\ (-1)^{J-M} \threej{J}{K}{J}{-M}{q}{M} \langle nSLJ || I^{(k)}_q || nSLJ \rangle.
\end{equation}
Then we reduce the matrix element further to one acting on total $L$:

\begin{equation}
    \label{eq:nSLJ}
    \langle nSLJ || I^{(k)}_q || nSLJ \rangle =\\ (-1)^{L+S+J+k} (2J+1) \sixj{L}{J}{S}{J}{L}{k} \langle nL || I^{(k)}_q || nL \rangle.
\end{equation}

Now, we would like to reduce $\langle nL || I^{(k)} || nL \rangle$ to a matrix element on the outer electron alone. However, since we are only interested in states where the inner electron is in $6s$ ($l_i=0$), we have the situation that $L = l_i + l_o = l_o$ and we can just replace $L$ with $l_o$ (the outer electron orbital angular momentum) in the reduced matrix element in Eq. (\ref{eq:nSLJ}). Therefore, the final result for the ponderomotive potential for alkaline Rydberg states of the form $msnl$ is the same as for alkali atoms [Eq. (\ref{eq:finalalkali})], evaluated with the \emph{total} angular momentum quantum numbers $SLJM$ instead of those for the Rydberg electron alone, $sljm$.

\subsection{Numerical evaluation of radial integrals}

To evaluate the potential numerically, the radial integrals must be performed. $f_{kq}(r)$ is computed using the gaussian solution to the paraxial wave equation \cite{Siegman1986Lasers_Book}. In the literature, several different approximations have been employed to find Rydberg wavefunctions for the computation of matrix elements using experimentally determined quantum defects, including Coulomb functions and numerical integration of the Schr\"odinger equation. In this work, we observe that the reduced matrix elements $\langle nl || I^{(k)}_q || nl \rangle$ depend only on the square modulus of the wavefunction and vary slowly with $n$ and $l$. Therefore, we compute the reduced matrix elements for integer $n$ using hydrogen radial wavefunctions $R_{nl}$, and interpolate between them to compute the effective matrix element for $n^* = n- \delta_{nSLJ}$, where $\delta_{nSLJ}$ is the measured quantum defect.

\subsection{Angular dependence}
Since the $n$ and $l$ dependence of the radial integrals is small, the variation of the trapping potential between nearby Rydberg states arises primarily from the angular factors in Eq. (\ref{eq:nSLJM}) - (\ref{eq:nSLJ}). These are tabulated for alkali and alkaline earth atoms in Table \ref{tab:angular}. In the alkali case, states of the same $j$ have the same angular factors and therefore approximately the same trapping potential \cite{Barredo2019}. In the alkaline case this is no longer true; however, the states $^1S_0$, $^3S_1$ and $^3P_0$ have the same (purely scalar) potential.

\begin{table}[h]
\begin{tabular}{ll|lll}
& Term & $k=0$ & $k=2$ & $k=4$ \\
\hline \hline
alkali & $^2S_{1/2}$&1&0&0\\ 
& $^2P_{1/2}$&1&0&0\\ 
&$^2P_{3/2}$&1&1/5&0\\ 
&$^2D_{3/2}$&1&1/5&0\\ 
&$^2D_{5/2}$&1&8/35&2/21\\ 
\hline
alkaline & $^1S_0$&1&0&0\\ 
& $^3S_1$&1&0&0\\ 
& $^1P_1$&1&2/5&0\\ 
& $^3P_0$&1&0&0\\ 
& $^3P_1$&1&-1/5&0\\ 
& $^3P_2$&1&1/5&0\\ 
& $^1D_2$&1&2/7&2/7\\ 
& $^3D_1$&1&1/5&0\\ 
& $^3D_2$&1&1/7&-4/21\\ 
& $^3D_3$&1&8/35&2/21\\ 
\end{tabular}
\caption{\label{tab:angular} Angular factors in $U_{r}$, expressed as the coefficient of the radial integral involving $f_{k0}$ for the $m=0$ state.}
\end{table}

\section{\label{Yb spec}Yb $^3S_1$ series spectroscopy}

\begin{figure}[h]
\includegraphics{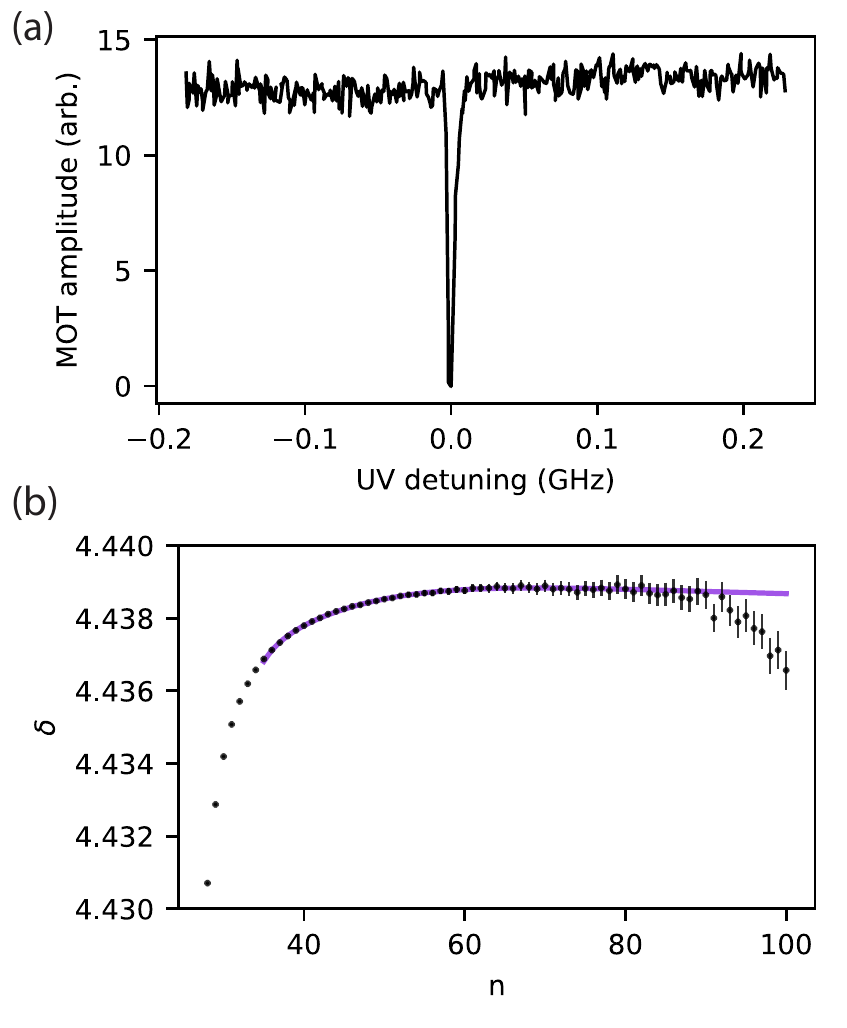}
\caption{\label{motspec}(a) Example spectrum for the \tripletS Rydberg line at $n=49$. (b) Measured quantum defects of the \tripletS Rydberg series vs. principal quantum number. Error bars show uncertainties in the defects from the 4 MHz uncertainty in the measured UV frequencies. The purple line is a fit to an extended Rydberg-Ritz model for $35<n<80$.}
\end{figure}

We measure the energies of the $^{174}$Yb $6sns$ \tripletS series using MOT depletion spectroscopy \cite{Couturier2019MeasurementAtoms, Hostetter2015MeasurementSpectroscopy}. First, we load a MOT on the \singletS to \tripletP transition in Yb. After compressing the MOT to $\sim$ 200 $\mu$m by ramping the power and detuning, we image the MOT on a camera while exposing it to a UV laser (about 5 mW in a $\sim$ 1 mm beam). Exciting atoms from \tripletP to a Rydberg state reduces the MOT fluorescence amplitude in the image. As detailed in the main text, we generate UV light at 308 nm by summing the output of a Ti:Sapphire laser at 1016 nm with a 1565 nm fiber laser and doubling the 616 nm output light in a resonant cavity. To perform high resolution spectroscopy, we change the frequency of the 1016 nm light and measure the frequency of the 616 nm output on a wavemeter with 2 MHz ($3\sigma$) accuracy (Toptica WS8-2). The wavemeter is calibrated with the $^{174}$Yb \singletS to \tripletP transition frequency (539386602.225 MHz). A typical spectrum is shown in Fig. \ref{motspec}a for $n=49$.

Table \ref{defectTable} presents the measured energies for the \tripletS series. The high resolution wavemeter gives an accuracy of 4 MHz in UV laser frequency. To determine the absolute energy of the levels, we add the $^{174}$Yb \singletS to \tripletP transition frequency, determined from the measured $^{171}$Yb  \singletS $F=1/2$ to \tripletP $F=3/2$ frequency (539390406.833 MHz) \cite{Nenadovi2016RelatedYb}, along with the isotope and hyperfine shifts from \cite{Pandey2009IsotopeYb}. By fitting the measured energies to $E_I - \frac{Ry}{(n-\delta)^2}$ in the region $60\leq n \leq 80$, we determine the $^{174}$Yb ionization energy to be $E_I = $ 50443.07074(4) cm$^{-1}$, within 10 MHz of the value reported in Ref. \cite{Lehec2018} ($Ry$ is the Rydberg constant for Yb, $109736.96959$ cm$^{-1}$).

The measured quantum defects for $n=28$ to $n=100$ are shown in Fig. \ref{motspec}b. To obtain an empirical model for the defects, we fit them to the extended Rydberg Ritz formula:
\begin{gather}
    \delta(n) = \delta_0 + \frac{\delta_2}{(n-\delta_0)^2} + \frac{\delta_4}{(n-\delta_0)^4} + ...
    \label{rydbergRitzDefect}
\end{gather}
The fit parameters for the region $35<n<80$ are summarized in Table. \ref{ritzFitParams}, and the fitted energies are within our experimental uncertainty in this range.
The measured defects are flat around $\delta = 4.439$ from $n=40$ to $n=80$.

\begin{table}[h]
\begin{tabular}{p{2.5cm}p{2.5cm}}
Fit Parameter  & Value      \\ \hline \hline
$\delta_0$ & 4.4382(2) \\
$\delta_2$ & 6(1)     \\
$\delta_4$ & $-1.8(4)\times10^4$ \\
$\delta_6$ & $1.8(5)\times10^7$ \\
$\delta_8$ & $-7(2)\times10^9$
\end{tabular}
\caption{\label{ritzFitParams} Parameters for the fit (Fig. \ref{motspec}b) to the extended Rydberg-Ritz model in Eq. (\ref{rydbergRitzDefect}), for $35<n<80$.}
\end{table}

At high $n$ ($n>85$), the measured energies (quantum defects) deviate significantly from the values extrapolated from the fit over $35<n<80$. The difference between the measured and extrapolated energies for $n=100$ is $+ 17 \pm 4$ MHz.

Several systematic effects may be responsible for this deviation. First, as noted in the text, we have intra-vacuum electrodes that allow us to zero the electric field in the region of the atoms. However, changing the potential on the electrodes results in an extremely long (10-30 minutes) settling time for the electric field seen by the Rydberg atoms, which we believe arises from significant field penetration into the glass and slow redistribution of charges there. In practice, this makes finding and maintaining a precise null difficult. The sign of the deviation (towards higher energy) is consistent with the measured sign of the quadratic Stark coefficient for the \tripletS series (positive), which is opposite that of Rb. The magnitude of the deviation is consistent with our estimated precision in setting the field null (which was carried out at $n=80$ before beginning data acquisition at $n=100$ and stepping down in $n$). However, the $n$ dependence of the deviation is closer to $n^{*12}$ than the expected $n^{*7}$ for Stark shifts. This could be attributed to slow settling of the electric fields over the measurement run (\emph{i.e.}, interpreting the apparent, approximate $n^{*12}$ dependence as a measurement artifact). We note that in more recent experiments (including those in the main text), the field settling time is shortened dramatically by illuminating the glass cell with a ~1 W UV LED at 365 nm (Ref. 27 of \cite{Levine2018High-FidelityQubits}). We believe the UV light ionizes charge traps in the glass cell, resulting in much faster equilibration of the charge distribution when the electrode potentials are changed. This allows field cancellation at the mV level. The resulting offset potentials are stable for weeks, which will allow more precise study of the high-$n$ spectroscopy in the future, although we were not able to repeat the high-precision spectroscopy because the necessary high-resolution wavemeter was no longer available to us.

A second possible systematic effect is van der Waals interactions, which would be consistent with the $n^{*12}$ scaling. However, in this case, one would expect nearly as much broadening as shifting, because of the random position of the atoms and varying density of the MOT. Empirically, we observe that the broadening is significantly less than the deviation (by about a factor of 3). Furthermore we estimate that the van der Waals interaction for this state is attractive (although it has not been measured), which is the wrong sign to explain the measured deviations. This estimate is based on the F\"orster defect for the $(n^3S_1, n^3S_1) \rightarrow (n^3P_2, (n-1)^3P_2)$ pair states of $-320$ MHz at $n=80$ ($\delta_{^3P_2}= 3.923$ \cite{Aymar1984}), which is significantly smaller in magnitude than the defect for other $P$ pair states and negative, and should therefore dominate the van der Waals interaction. This is in contrast to $^1S_0$, which is predicted to have a small, repulsive interaction \cite{Vaillant2012Long-rangeYtterbium}. 

A final possibility is the presence of a series perturber at or above the first ionization limit. Further experiments with more carefully controlled systematics at high-$n$ will be necessary to disentangle these possibilities.

\begin{table}[h]
\begin{tabular}{p{1.5cm}p{2.5cm}|p{1.5cm}p{2.5cm}}
$n$ & Energy (cm$^{-1}$) & $n$ & Energy (cm$^{-1}$) \\
\hline \hline
28 & 50245.5285 & 64 & 50412.1374 \\
29 & 50261.2497 & 65 & 50413.1505 \\
30 & 50275.1772 & 66 & 50414.1147 \\
31 & 50287.5688 & 67 & 50415.0329 \\
32 & 50298.6401 & 68 & 50415.9083 \\
33 & 50308.5712 & 69 & 50416.7432 \\
34 & 50317.5130 & 70 & 50417.5402 \\
35 & 50325.5925 & 71 & 50418.3016 \\
36 & 50332.9169 & 72 & 50419.0294 \\
37 & 50339.5773 & 73 & 50419.7256 \\
38 & 50345.6515 & 74 & 50420.3921 \\
39 & 50351.2064 & 75 & 50421.0303 \\
40 & 50356.2995 & 76 & 50421.6420 \\
41 & 50360.9806 & 77 & 50422.2285 \\
42 & 50365.2929 & 78 & 50422.7914 \\
43 & 50369.2741 & 79 & 50423.3316 \\
44 & 50372.9574 & 80 & 50423.8507 \\
45 & 50376.3717 & 81 & 50424.3495 \\
46 & 50379.5425 & 82 & 50424.8291 \\
47 & 50382.4925 & 83 & 50425.2906 \\
48 & 50385.2417 & 84 & 50425.7348 \\
49 & 50387.8079 & 85 & 50426.1625 \\
50 & 50390.2071 & 86 & 50426.5745 \\
51 & 50392.4533 & 87 & 50426.9718 \\
52 & 50394.5593 & 88 & 50427.3548 \\
53 & 50396.5366 & 89 & 50427.7242 \\
54 & 50398.3955 & 90 & 50428.0809 \\
55 & 50400.1451 & 91 & 50428.4254 \\
56 & 50401.7940 & 92 & 50428.7578 \\
57 & 50403.3496 & 93 & 50429.0794 \\
58 & 50404.8190 & 94 & 50429.3902 \\
59 & 50406.2082 & 95 & 50429.6906 \\
60 & 50407.5232 & 96 & 50429.9813 \\
61 & 50408.7690 & 97 & 50430.2627 \\
62 & 50409.9505 & 98 & 50430.5352 \\
63 & 50411.0720 & 99 & 50430.7988 \\
64 & 50412.1374 & 100 & 50431.0545 \\
\hline \hline
\end{tabular}
\caption{\label{defectTable}Measured energies of the $^{174}$Yb $6sns$ \tripletS Rydberg series from $n=28$ to $n=100$.}
\end{table}

\section{Trap-induced loss mechanisms}
\label{trap loss}

\begin{table}[h]
\centering
\begin{tabular}{p{3cm}p{2cm}}
state         & Lifetime ($\mu$s) \\ \hline \hline
$n = 74\; ^3P_2$ &  83(5)           \\ \hline
$n = 74\; ^3P_0$ &  14(4)$^*$          \\ \hline
$n = 73\; ^1D_2$ & 75(18)           \\
$n = 83\; ^1D_2$ & 59(3)           \\
$n = 90\; ^1D_2$ & 60(3) \\ \hline
$n = 70\; ^3S_1$ & 85(4)      \\
$n = 75\; ^3S_1$ & 105(3) \\
$n = 92\; ^3S_1$ & 42(2)\\
\end{tabular}
\caption{Summary table of power dependent lifetime studies for various Rydberg states, showing the extrapolated lifetimes with no trap-induced losses. $^*$We do not have power dependence data to quote the extrapolated lifetime for \tripletPzero, but give the value for 9 mW trap power.}
\label{tab:ltvspwr}
\end{table}

\subsection{Auto-ionization}
The polarizability of the Yb$^+$ ion core primarily arises from the $6s$ to $6p_{1/2}$ and $6p_{3/2}$ transitions at 369 and 329 nm, respectively. For an optically trapped Yb$^+$ ion, the finite lifetime of these states ($1/\Gamma = 8$ ns for $6p_{1/2}$, \cite{Hayes2009MeasurementYb+}) would give rise to photon scattering at a rate $\Gamma \Omega^2/\Delta^2 = \Gamma U_0/\Delta$, where $\Omega$ and $\Delta$ are the Rabi frequency and detuning of the trapping laser, and $U_0$ is the trap depth. In the case of a trapped Yb$^0$ Rydberg state, the core transitions are between $6snl$ and $6p_jnl$ states (where $n,l$ are the quantum numbers of the outer Rydberg electron). Since the $6p_jnl$ states are above the ionization limit of the $6snl$ series, they can rapidly auto-ionize. In an auto-ionization event, the core electron is de-excited by ejecting the Rydberg electron. The rate of this process decreases as $1/n^{*3}$, since the interaction between the Rydberg and core electron decreases as the former moves farther out. The rate of these events is the same as photon scattering for a Yb$^{+}$ ion, with the linewidth $\Gamma$ replaced by the auto-ionizing linewidth $\gamma' n^{*-3}$, according to the isolated core electron approximation \cite{Cooke1978DoublySr}.

The auto-ionization rates have been measured for certain $5p_jns$ states in Sr \cite{Xu1986} and $6p_jns$ states in Ba (cited in Ref. 21 of \cite{Xu1986}) and Yb \cite{Xu1994}. All are within the range of $\gamma' = 2 \pi \times 10^{14}$ - $2 \pi \times 10^{15}$ s$^{-1}$. The rate for the Yb $6p_{3/2}ns$ series has not been measured to the best of our knowledge, but for concreteness we take the measured value for the $6p_{1/2}ns$ series of $\gamma' = 1.2 \times 10^{15}$ s$^{-1}$ \cite{Xu1994}, and assume the $6p_{3/2}ns$ rate is 2 times higher, which is approximately the case for Sr and Ba. From this, we can estimate an auto-ionization rate of about $n^{*-3}\: 58 \times 10^6$ s$^{-1}$ at typical trap powers (9 mW). This corresponds to a 7 ms lifetime at $n=75$, which is several orders of magnitude slower than the \tripletS lifetimes we observe. The rate should be even smaller for higher $l$ states, since the overlap with the core decreases for high $l$. This is supported by measurements \cite{Cooke1978DoublySr, Fields2018AutoionizationStates} for $l<7$ and theory predicting an $l^{-5}$ scaling of the autoionization rate for high $l$ \cite{Cooke1979CalculationStates}.

\subsection{Photo-ionization}

\begin{figure}[h]
\includegraphics{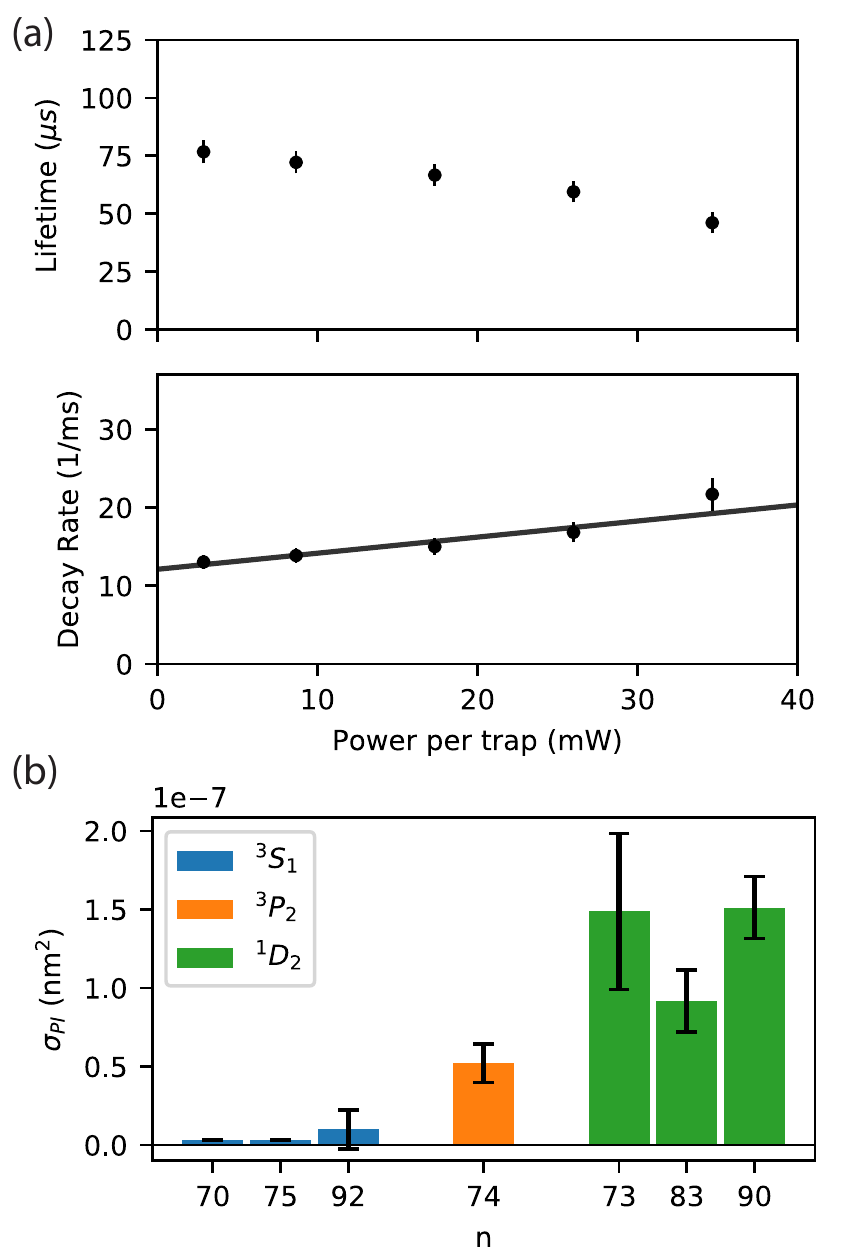}
\caption{\label{fig:photoionization}(a) Rydberg lifetimes vs. trap power for the $n=74 ^3P_2$ state. The lower plot shows the corresponding decay rates and a linear fit, the slope of which determines the photoionization cross section $\sigma_{PI}$. (b) Measured photoionization cross sections for $^3S_1$, $^3P_2$, and $^1D_2$ states at different $n$. We see no evidence of photoionization for $S$ states.} 
\end{figure}

Another loss mechanism is direct excitation of the Rydberg electron into continuum states, termed photoionization \cite{Gallagher1994, Saffman2005}. Calculating this process requires computing matrix elements between bound and free Rydberg states, which requires either model potentials or extrapolation of measured bound state quantum defects into the continuum \cite{Burgess1958}. The extrapolation quantum defects based on the Rydberg-Ritz model has been studied for alkali atoms \cite{Burgess1958}; however, given the very strong $n$-dependence near the ionization threshold observed for the Yb $^3P_0$ and $^3P_1$ states \cite{Aymar1984}, a multi-channel quantum defect model would be necessary for a precise calculation of photo-ionization rates for Yb. Furthermore, it is possible to have perturbers above the ionization threshold that would not be evident from the bound state quantum defects.

 We instead determine the photoionization rates for a few $S$, $P$, and $D$ states by measuring the Rydberg state lifetime as a function of trap power. Here, we assume that all power-dependent loss results from photoionization, since the predicted autoionization rates are very small. Fig. \ref{fig:photoionization}a shows an example measurement for the $n=74$ \tripletPtwo state with the corresponding decay rates and a fit to $\Gamma = \Gamma_0 + \gamma_{PI} P$, where $P$ is the trap power. Extrapolating the decay rates to zero trap power gives an estimate for the natural Rydberg state lifetimes $1/\Gamma_0$, which are summarized in Table \ref{tab:ltvspwr}. Fig. \ref{fig:photoionization}b shows the photoionization cross sections for the measured states, related to the decay rates by $\gamma_{PI}P = \sigma_{PI} I /\hbar \omega$. The measured photoionization cross sections are consistent with zero for the \tripletS series, but the rates are non-negligible for the \tripletPtwo and $^1D_2$ states, resulting in a 15-30\% reduction of the trapped lifetimes at typical trap powers. The observed magnitude and low-$L$ dependence of the cross section is similar to previous calculations for Rb \cite{Saffman2005}. As with auto-ionization, the photoionization rate should decrease for high-$L$ Rydberg states, in this case exponentially with $L$ \cite{Nguyen2018}.

\section{\tripletS Lifetime at high $n$}
Given that we observe no trap-induced losses for the \tripletS series, the measured decrease in the \tripletS lifetimes at high $n$ is surprising. For alkali atoms, the finite Rydberg state lifetime is a combination of spontaneous emission to much lower-$n$ Rydberg states, and blackbody-induced transitions between nearby Rydberg levels \cite{Saffman2010}. These rates scale as $n^{*-3}$ and $n^{*-2}$ respectively. Lifetimes of greater than 250 $\mu$s have been measured for Rb Rydberg states at $n=85$ at room temperature \cite{Archimi2019MeasurementsLevels}. In alkaline earth atoms, an additional loss mechanism is present: configuration interactions can mix Rydberg series attached to different ionization thresholds associated with different core ion states, which results in admixtures of fast-decaying low-$n$ states into the high-$n$ states. Rydberg state energies and lifetimes can be highly irregular near a perturbing resonance. Far from a resonance, however (e.g., at high $n$), the impact of the perturber is a constant factor reduction of the Rydberg series radiative lifetime \cite{Vaillant2014}. Perturbers are almost certainly responsible for the measured, short lifetime of the $^3P_0$ series (Table \ref{tab:ltvspwr}), as this series is known to be extremely strongly perturbed \cite{Aymar1984Three-stepYtterbium}.

Whether series perturbations are responsible for the drop in \tripletS lifetime at high-$n$ is unclear. While the low-$n$ \tripletS spectrum (Fig. \ref{motspec}) appears relatively unperturbed (compared to \tripletPzero, for example, where the quantum defect varies by nearly 0.5 between $n=30$ and $n=80$ \cite{Aymar1984Three-stepYtterbium}), the deviation from constant $\delta$ at high-$n$ could reflect an above-threshold perturber, or a technical artifact in the measurement, as discussed in section \ref{Yb spec}. Another possibility is high-frequency electric field noise conducted to the atoms by a set of intra-vacuum electrodes around the atoms. We have simulated the local density of states in the center of the electrodes at microwave frequencies and find low-$Q$ resonances giving rise to Purcell enhancement factors of 2-3 at 7 and 15 GHz, where the strongest transitions lie for $n=65-100$, which could enhance the blackbody transition rate by a similar factor or enhance the coupling of noise at these frequencies. This would also reduce certain high-$n$ lifetimes for other series as well, but we do not have sufficient data on the $n$-dependence of the $P$ and $D$ series to confirm this. At present, we are unable to experimentally distinguish whether the high-$n$ lifetimes are intrinsic or limited by technical effects.

Lastly, we note that the theoretical lifetime can in principle be extracted from MQDT analysis of the Rydberg spectrum. This has been attempted in some detail for Strontium \cite{Vaillant2014}. In the case of Yb, this is hampered by the absence of complete spectroscopic data, the larger number of perturbers that result from low-lying $f$-shell excitations in the core (e.g., states of the form $4f^{13}5d6s$), and the fact that the lifetimes of the open $f$-shell states are not known in many cases. The lifetimes presented in this paper are, to the best of our knowledge, the first reported (or predicted) lifetimes for high-$n$ Yb Rydberg states. Together with additional spectroscopic data, this will aid future analyses of the complete Yb Rydberg spectrum.

\bibliographystyle{apsrev4-2}
\bibliography{finalbib}